\documentclass[twocolumn, showpacs,preprintnumbers,amsmath,amssymb]{revtex4}
\usepackage{amssymb}
\usepackage{xcolor}
\usepackage{graphicx}% Include figure files
\usepackage{dcolumn}% Align table columns on decimal point
\usepackage{bm}% bold math
\usepackage{multirow}
\usepackage{booktabs}
\usepackage{CJK}
\usepackage{natbib}
\usepackage{soul}
\usepackage{amssymb}
\usepackage{amsmath}	% Advanced maths commands
\usepackage{color}
\usepackage{cases} 

\newcommand{\Vmat}{{\boldmath $\cal V$}}
\newcommand{\Kmat}{{\boldmath $\cal K$}}

\newcommand{\numat}{\mbox{\boldmath $\nu$}}

\newcommand{\Smat}{\mbox{\boldmath $\cal S$}}
\newcommand{\SSmat}{\mbox{\boldmath $S$}}
\newcommand{\Xmat}{\mbox{\boldmath $X$}}

\begin{document}
\preprint{APS/123-QED}

%\title{ Optical shielding of ultracold K-Cs binary collision}
\title{Dissociative recombination and rotational transitions of D$_2^{+}$ in collisions with slow electrons}
\begin{CJK*}{GBK}{song}\end{CJK*}

\author{\begin{CJK*}{GBK}{song}M. D. Ep\'ee Ep\'ee$^{1}$\end{CJK*}}%
\author{\begin{CJK*}{GBK}{song}O. Motapon$^{1,2}$\end{CJK*}}
\author{\begin{CJK*}{GBK}{song}N. Pop$^{3}$\end{CJK*}}
\author{\begin{CJK*}{GBK}{song}F. Iacob$^{4}$\end{CJK*}}
\author{\begin{CJK*}{GBK}{song}E. Roueff$^{5}$\end{CJK*}}
\author{\begin{CJK*}{GBK}{song}I. F. Schneider$^{6,7}$\end{CJK*}}
\author{\begin{CJK*}{GBK}{song}J. Zs Mezei$^{6,8}$\end{CJK*}}
\email[]{mezei.zsolt@atomki.hu}
\affiliation{$^{1}$Department of Physics, Faculty of Sciences, University of Douala, P. O. Box 24157, Douala, Cameroon}%
\affiliation{$^{2}$Faculty of Science, University of Maroua, P. O. Box 814, Maroua, Cameroon}%
\affiliation{$^{3}$Department of Physical Foundation of Engineering, University Politechnica of Timisoara, 300223, Timisoara, Romania}%
\affiliation{$^{4}$Physics Faculty, West University of Timisoara, 300223,Timisoara, Romania}%
\affiliation{$^{5}$LERMA CNRS-UMR8112, Observatoire de Paris, Universit\'e PSL, F-92190, Meudon, France}%
\affiliation{$^{6}$LOMC CNRS-UMR6294, Universit\'e le Havre Normandie, F-76058 Le Havre, France}%
\affiliation{$^{7}$LAC CNRS-FRE2038, Universit\'e Paris-Saclay, F-91405 Orsay, France}%
\affiliation{$^{8}$Institute for Nuclear Research (ATOMKI), H-4001 Debrecen, Hungary}%
\date{\today}

\begin{abstract}
Rate coefficients for dissociative recombination and state-to-state rotational transitions of the D$_{2}^{+}$ ion induced by collisions with very low-energy electrons have been reported following our previous studies on HD$^{+}$ and H$_{2}^{+}$ \cite{motapon2014,Epee2015}. 
The same molecular structure data sets, excitations ($N_{i}^{+} \rightarrow$  $N_{f}^{+}=N_{i}^{+}+2$ for $N_{i}^{+}=0$ to $10$) and de-excitations ($N_{i}^{+}$ $\rightarrow$ $N_{f}^{+}=N_{i}^{+}-2$, for $N_{i}^{+}=2$ to $10$) were used for collision energies ranging from $0.01$ meV to $0.3$ eV. Isotopic effects for dissociative recombination and rotational transitions of the vibrationally relaxed targets are presented.
\end{abstract}

%\keywords{coupled-channel, optical shielding, KCs}%Use showkeys class option if keyword

\maketitle

\section{Introduction}

Among many "cold" ionised environments, in the diffuse interstellar media and planetary atmospheres, electrons are presumed to be one of the most important exciting species for molecular cations. Recent studies based on astrophysical observations and calculations regarding diatomic and polyatomic molecular charged species~\citep{faure01,faure02,faure03,faure04,slava01} show that the cross sections of electron impact rotational transitions of molecular cations significantly exceed those obtained by atomic and/or molecular impact.
 
At very low electron collision energy,  the electron-impact induced rotational transitions of the vibrationally relaxed molecular cations:

\begin{equation}
\label{eq:SEC} 
{\text AB}^{+}(N_{i}^{+},v_{i}^{+}=0)+e^{-}(\varepsilon) 
\longrightarrow 
{\text AB}^{+}(N_{f}^{+},v_{f}^{+}=0)+e^{-}({\varepsilon}')
\end{equation}

\noindent are in strong competition with the dissociative recombination:
\begin{equation}
\label{eq:DR} 
{\text AB}^{+}(N_{i}^{+},v_{i}^{+}=0)+e^{-}(\varepsilon) 
\longrightarrow 
{\text A} + {\text B},
\end{equation}
\noindent Here $N_{i}^{+}/N_{f}^{+} $ and $v_{i}^{+}/v_{f}^{+} $ stand for the initial/final rotational and vibrational quantum numbers of the cation and $\varepsilon /\varepsilon'$ the kinetic energy of the incident/scattered electrons. 

 Recently, in the Le Havre group, several studies were performed on the electron-induced reactions  of H$_{2}^{+}$ and HD$^{+}$~\citep{motapon2008,waffeutamo2011,chakrabarti2013,motapon2014,Epee2015,djuissi2020}. State-to-state cross sections and rate coefficients were reported for ro-vibrational transitions including inelastic collisions (IC, $N_{i}^{+}  < N_{f}^{+}$ and/or $v_{i}^{+}<v_{f}^{+}$), super-elastic collisions (SEC, $N_{i}^{+}>N_{f}^{+}$ and/or $v_{i}^{+}>v_{f}^{+}$), dissociative recombination (DR) and - at high collision energies - dissociative excitation (DE).  
 
 The calculations were performed within the framework of our stepwise multichannel quantum defect theory~\citep{annick1980, mezei2019}  (and references therein). 

These results were compared with experimental results obtained for HD$^{+}$ and  H$_{2}^{+}$ on different merged beam~\citep{auerbach1977,hus1988} and storage ring experiments performed at the Heidelberg Test Storage Ring \citep{shafir,schwalm,krohn2000,krohn2001} and the Aarhus Astrid Storage Ring~\citep{andersen1997}. The calculated cross sections and rate coefficients agree satisfactorily with the measured ones for both DR and ro-vibrational excitation processes.

It is also interesting to report that simple deuterium chemistry has been suggested to take place in the primordial conditions, where essentially only Hydrogen, Deuterium and Helium elements are present \citep{lepp:84}. The presence of HD, with its small permanent dipole moment, may indeed contribute to the cooling  of the medium. Several groups have  explicitly introduced coupled Hydrogen/Deuterium/ Helium chemistry as emphasized in the review paper of \cite{galli:13}. However, to our knowledge, only \cite{gay:11} have explicitly introduced multiple deuterated compounds in this context, including D$_2^+$, HD$_2^+$ and D$_3^+$ and conclude their paper by emphasizing the large uncertainties present in their deuterium chemistry and claim for additional studies on the topic. As an example, they suggest that the dissociative recombination rate coefficient of D$_2^+ $ is equal to $1.2 \times 10^{-8} (\frac{T}{300})^{-0.4}$, which is equal to their estimate of the H$_2^+$ dissociative recombination rate coefficient. The aim of the present study is to overcome that assumption and to explicitly consider the different nuclear effects for the dissociative recombination of the heavy D$_2^+$ molecular ion, as a natural extension of previous studies on H$_2^+$ and HD$^+$~\citep{motapon2014,Epee2015}. 

The paper is organised as follows. In section \ref{sec:theory} we briefly describe our theoretical approach. Rate coefficients and their comparison with previous results are presented in section~\ref{sec:results}, and the conclusions follow in section~\ref{sec:conclusions}.

\section{Theoretical Method}{\label{sec:theory}}

The efficiency of our theoretical method in modelling the electron/diatomic cation collisions, based on the {\it stepwise} Multichannel Quantum Defect Theory (MQDT), has been proved in many previous studies on different species, including H$_2^+$ and its isotopologues~\cite{motapon2008,waffeutamo2011,chakrabarti2013,motapon2014,Epee2015,djuissi2020}, CH$^+$~\cite{mezei2019}, SH$^+$~\cite{Kashinski2017},  etc.  
The general ideas of our approach were already presented in detail many times, see for example~\cite{mezei2019} and, therefore, here we restrict ourselves to its major steps.

The reactions~(\ref{eq:SEC}) and (\ref{eq:DR}) involve \textit{ionization} channels - characterising the scattering of an electron on the target cation - and \textit{dissociation} channels - relating to atom-atom scattering. The mixing of these channels results in quantum interference of the \textit{direct} mechanism - in which the capture takes place into a doubly excited dissociative state of the neutral system - with the \textit{indirect} one - in which the capture occurs via a Rydberg bound state of the molecule belonging to a \textit{closed} channel, this state being predissociated by the dissociative one. 
In both mechanisms the autoionization - based on the existence of \textit{open}  ionization channels - is in competition with the predissociation, and can lead to the excitation or to the de-excitation of the cation.

More specifically, each of the ionization channels, built by adding an electron to the D$_2^+$ ion in its  ground electronic state  
$X$ ${^2}\Sigma{_g^+}$ in a particular vibrational level, interacts with all the dissociation exit channels (Rydberg-valence interaction) for all the relevant symmetries ($^1\Sigma_{g}^{+}$, $^1\Pi_{g}$, $^1\Delta_{g}$, $^3\Sigma_{g}^{+}$, $^3\Pi_{g}$, $^3\Delta_{g}$, $^3\Sigma_{u}^{+}$, and $^3\Pi_{u}$). 
Depending on the total energy of the system these ionization channels can be {\it open} - either as entrance channels, describing the incident electron colliding the ion in its ground electronic state, or  exit channels, describing the auto-ionization, i.e. resonant elastic scattering, ro-vibrational excitation and de-excitation  - or {\it closed} - describing the resonant temporary captures into Rydberg states. 

The MQDT treatment of DR and rotational transitions requires the {\it a priori} knowledge of the potential energy curves (PECs) of the ion ground state and the relevant doubly excited, dissociative states of the neutral molecule, as well as for the Rydberg series of mono-excited states represented by the quantum defects. The driving forces  of the the recombination and excitation processes are the electronic couplings that connects the dissociative and ionization continua. 

At low collision energies, besides the type of the crossing of the neutral states with the ground ion state (favourable or less favourable crossing), the rotational couplings among the neutral states with different symmetries are of key importance~\citep{motapon2014}.
In our calculations the $^1\Sigma_{g}^{+}$ symmetry rotationally couples to $^1\Pi_{g}$ and $^1\Delta_{g}$, $^3\Sigma_{g}^{+}$ couples to  $^3\Pi_{g}$ and $^3\Delta_{g}$ and, finally, $^3\Sigma_{u}^{+}$ couples to $^3\Pi_{u}$. For the remaining symmetries the electronic couplings are for at least two orders of magnitude smaller so they can be neglected. Most of these data were extracted from {\it ab initio} molecular structure calculations of Wolniewicz {\it et al}~\citep{kolos1969,wolniewicz1994,orlikowski1999,staszewka2002}, completed by R-matrix calculations of~\citep{tennyson1996} and~\citep{telmini2003}. 
For each of the symmetries involved, only the lowest dissociative states are considered since they are the relevant ones in low-energy collisions. As for the ionization channels, the partial waves considered for the incident electron were $s$ and $d$ for the $^1\Sigma_{g}^{+}$ states, $d$ for  $^1\Pi_{g}$, $^1\Delta_{g}$, $^3\Sigma_{g}^{+}$, $^3\Pi_{g}$ and $^3\Delta_{g}$, and $p$ waves for $^3\Sigma_{u}^{+}$, and $^3\Pi_{u}$. 

 The first step in our approach is to build the {\it interaction matrix} {\Vmat} that drives the collision, whose elements quantify the couplings between  the different channels - ionization and dissociation ones. 

Once the {\Vmat}-matrix is built, we construct the short-range reaction matrix {\Kmat} of the collision, as a second order perturbative solution of the Lippmann-Schwinger equation. The diagonalized version of the {\Kmat}-matrix (in the eigenchannel representation) whose eigenvalues are expressed in terms of long range phase-shifts of the eigenfunctions, together with the vibronic couplings between the ionization channels, serve for the building of the frame transformation matrices. 

Applying a Cayley transformation on these latter matrices we can set up the generalized scattering matrix {\Xmat}.
 The Seaton's method of 'eliminating' the closed channels~\cite{Seaton1983} is then employed, resulting in the physical scattering matrix {\Smat}:
\begin{equation}
\SSmat = \Xmat_{oo}-\Xmat_{oc}\frac{1}{\Xmat_{cc}-\exp({\rm -i 2 \pi} \numat)} \Xmat_{co}\,,
\label{eq:elimination}
\end{equation}
relying on the block-matrices involving open ({\Xmat$_{oo}$}), open and closed ({\Xmat$_{oc}$} and {\Xmat$_{co}$}) and closed (\Xmat$_{cc}$) channels. The diagonal matrix {\numat} in the denominator of equation (\ref{eq:elimination}) contains the effective quantum numbers corresponding  to the  the vibrational thresholds of the closed ionisation channels at given total energy of the system.

\begin{figure*}
		\includegraphics[width=\columnwidth]{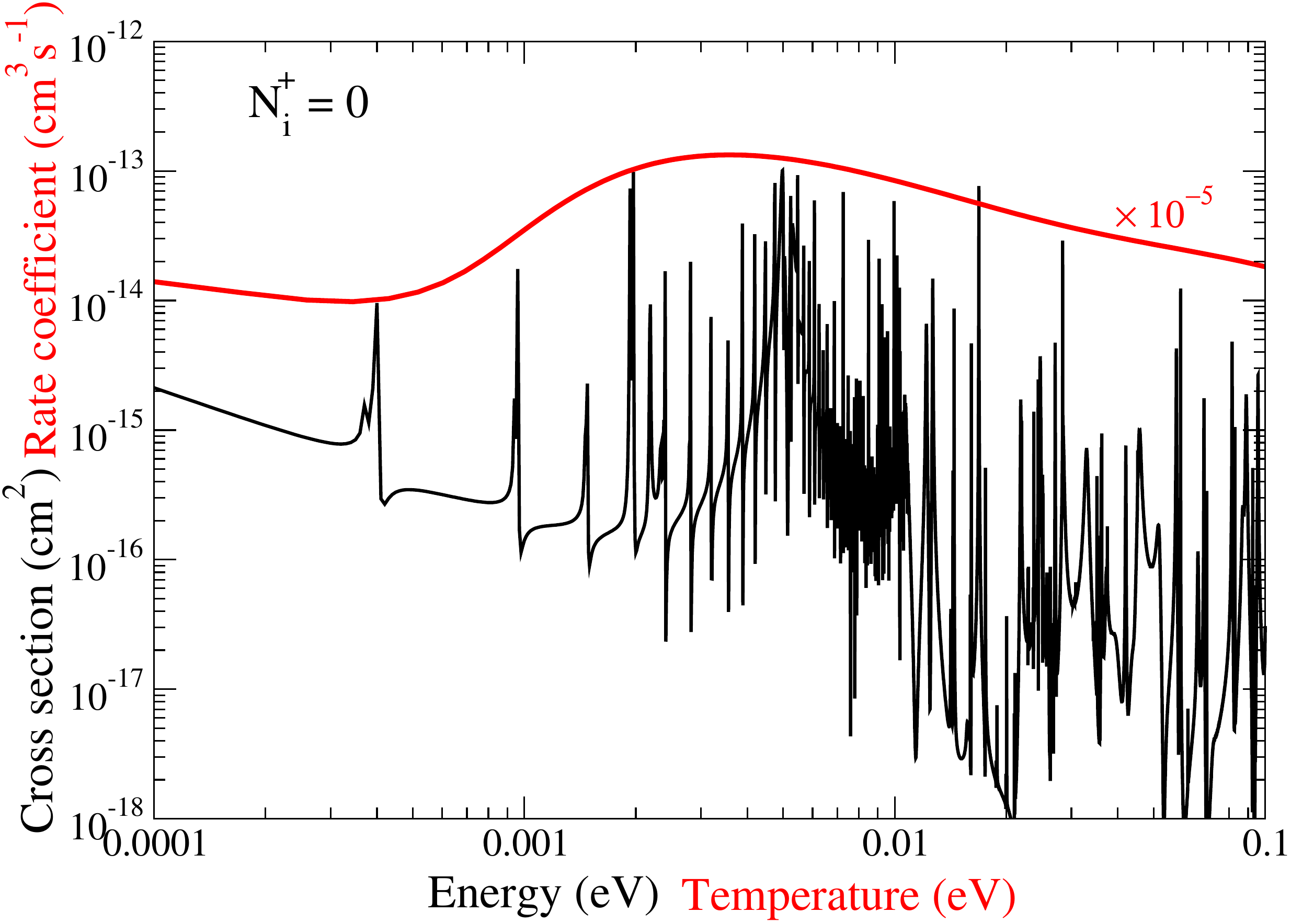}
		\includegraphics[width=\columnwidth]{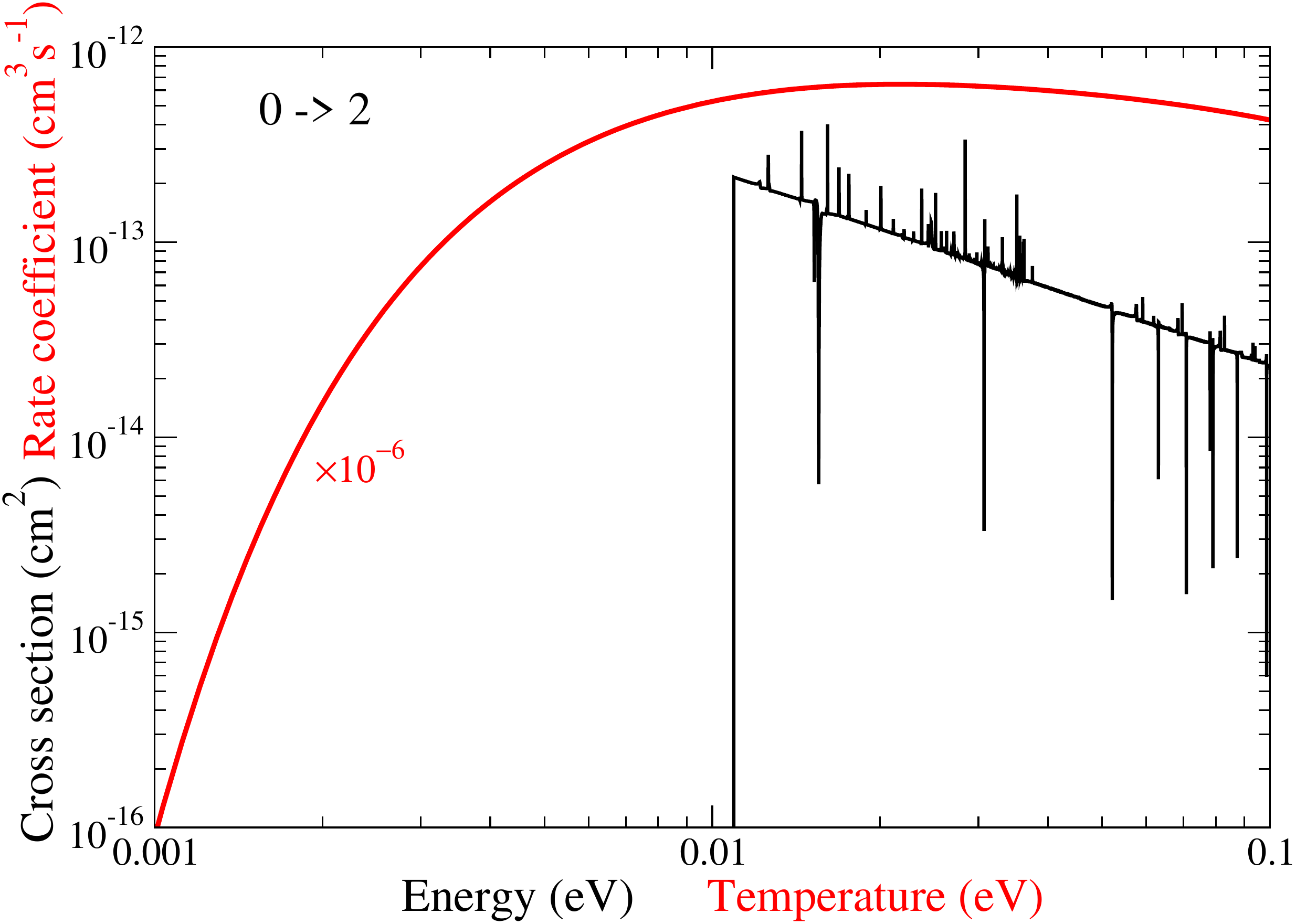}

    \caption{Cross section (black thin line) and thermal rate coefficient (red thick line) of dissociative recombination (left panel) and rotational excitation (right panel) of  ground state D$_{2}^{+}$ (X $^{2}\Sigma_{g}^{+}$, $N_i^+=0, v_i^+=0)$. The thermal rate coefficients are scaled by 10$^{-5}$ and 10$^{-6}$ respectively.
    }
    \label{fig:xs}
\end{figure*}

Finally, the global cross section for the dissociative recombination and for the rotational transitions - Rotational Excitation/de-Excitation (RE/RdE) and resonant elastic scattering of a vibrationally-relaxed ion reads as:
 
\begin{equation}
\begin{split}
\sigma _{diss \leftarrow N_{i}^{+}}  &= \sum_{\Lambda,sym} \frac{\pi}{4\varepsilon} \rho^{(sym,\Lambda)} \sum_{N} \frac{2N+1}{2N_{i}^{+}+1} \times \\
		&\times\sum_{l,j}\mid S^{(sym,\Lambda,N)}_{d_{j},N_{i}^{+}l}\mid^2,\label{eqDR2} 
\end{split}
\end{equation}		
and
\begin{equation}
\begin{split}
\sigma _{N_{f}^{+} \leftarrow N_{i}^{+}}  &= \sum_{\Lambda,sym} 
		\frac{\pi}{4\varepsilon} \rho^{(sym,\Lambda)} \sum_{N} \frac{2N+1}{2N_{i}^{+}+1}\times\\
		&\times\sum_{l,l'}\mid S_{N_{f}^{+}l',N_{i}^{+}l}^{(sym,\Lambda,N)} -\delta_{N_{i}^{+}N_{f}^{+}}\delta_{l'l}\mid^2 ,\label{eqVE_VdE2}
\end{split}
\end{equation}
 where $sym$ is referring to the inversion symmetry  - gerade/ungerade - and to the spin quantum number of the neutral system,   $N$ stands for its total rotational quantum number (for more details see Table 1. from~\cite{schneider1997}), $N_{i}^{+}/N_{f}^{+} $ denote the initial/final rotational quantum number of the cation and $\rho^{(sym,\Lambda)}$ is the ratio between the multiplicities of the neutral system and of the ion.

The thermal rate coefficients have been obtained by the convolution of the cross section with the Maxwellian isotropic energy distribution function for the free electrons:
\begin{equation}\label{rate}
%\alpha(T)=\frac{8\pi m}{{(2\pi mkT)}^{3/2}}\int_{0}^{+\infty}\sigma(\varepsilon)\varepsilon\exp(-\varepsilon/kT)d\varepsilon,
\alpha(T)=\frac{2}{kT}\sqrt{\frac{2}{\pi mkT}}\int_{0}^{+\infty}\sigma(\varepsilon)\varepsilon\exp(-\varepsilon/kT)d\varepsilon,
\end{equation}
 where $m$ is the mass of the electron, $T$ stands for the temperature and $k$ is the Boltzmann constant.

\section{Results and discussions}{\label{sec:results}}

Applying the stepwise MQDT method outlined in the previous section we have calculated the dissociative recombination (eq.~(\ref{eqDR2})), and rotational transition (excitation and de-excitation) (eq.~(\ref{eqVE_VdE2})) cross sections of D$_{2}^{+}$ for its lowest $11$ ($N_{i}^{+}=0-10$) rotational levels of its ground vibrational level ($v_{i}^{+}=v_{f}^{+}=0$). The electron impact collision energies range between $0.01$ and  $300$ meV. Convoluting these cross sections conform eq.~(\ref{rate}), we obtain the DR, RE and RdE thermal rate coefficients for electron temperatures ranging between $10$ and $1000$ K.

\begin{figure}
		\includegraphics[width=\columnwidth]{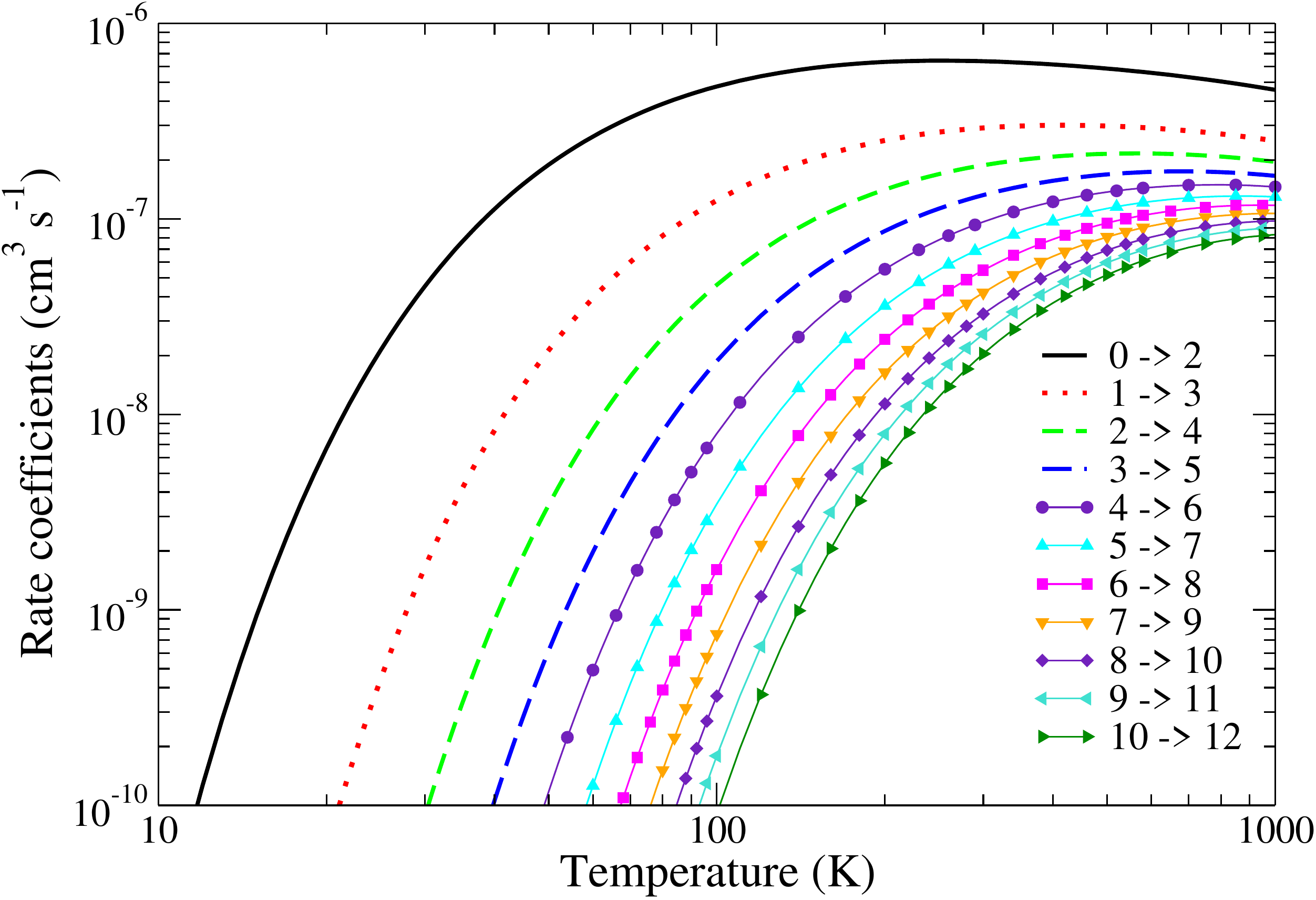}
    \caption{Maxwell rate coefficients for the rotational excitation $N_{i}^{+} \rightarrow N_f^{+}=N_{i}^{+}+2$, with $N_{i}^{+}=0$ to $10$ of the vibrationally relaxed $(v_i^+=0)$ D$_{2}^{+}$ (X $^{2}\Sigma_{g}^{+})$.}
    \label{fig:r_IC_iso}
\end{figure}

In Figure~\ref{fig:xs} we show a typical behaviour of the DR ($N_{i}^{+}=0$, left panel) and of the RE ($N_i^+=0\rightarrow N_f^+=2$, right panel) cross sections and their  thermal rate coefficients. In black we represent the cross sections as function of collision energy while in red we give the scaled rate coefficients as function of the electron temperature, the scaling factors being also given. In the left panel one can notice how the cumulation of the narrow constructive Rydberg resonances at about 6 meV will produce a maximum in the shape of the DR rate coefficient. The right panel of the same figure gives us the general form of the RE rate coefficient, where the sharp threshold present in the cross section is averaged out into a monotonically increasing function. 																																													
Figure \ref{fig:r_IC_iso} shows the rate coefficient for the consecutive $N_{i}^{+} \rightarrow N_{i}^{+}+2$ excitations - allowed by the selection rules -  for $N_{i}^{+}=0-10$ initial rotational quantum numbers. One can notice that their magnitudes are monotonically decreasing as $N_{i}^{+}$ is increased.

\begin{figure}
		\includegraphics[width=\columnwidth]{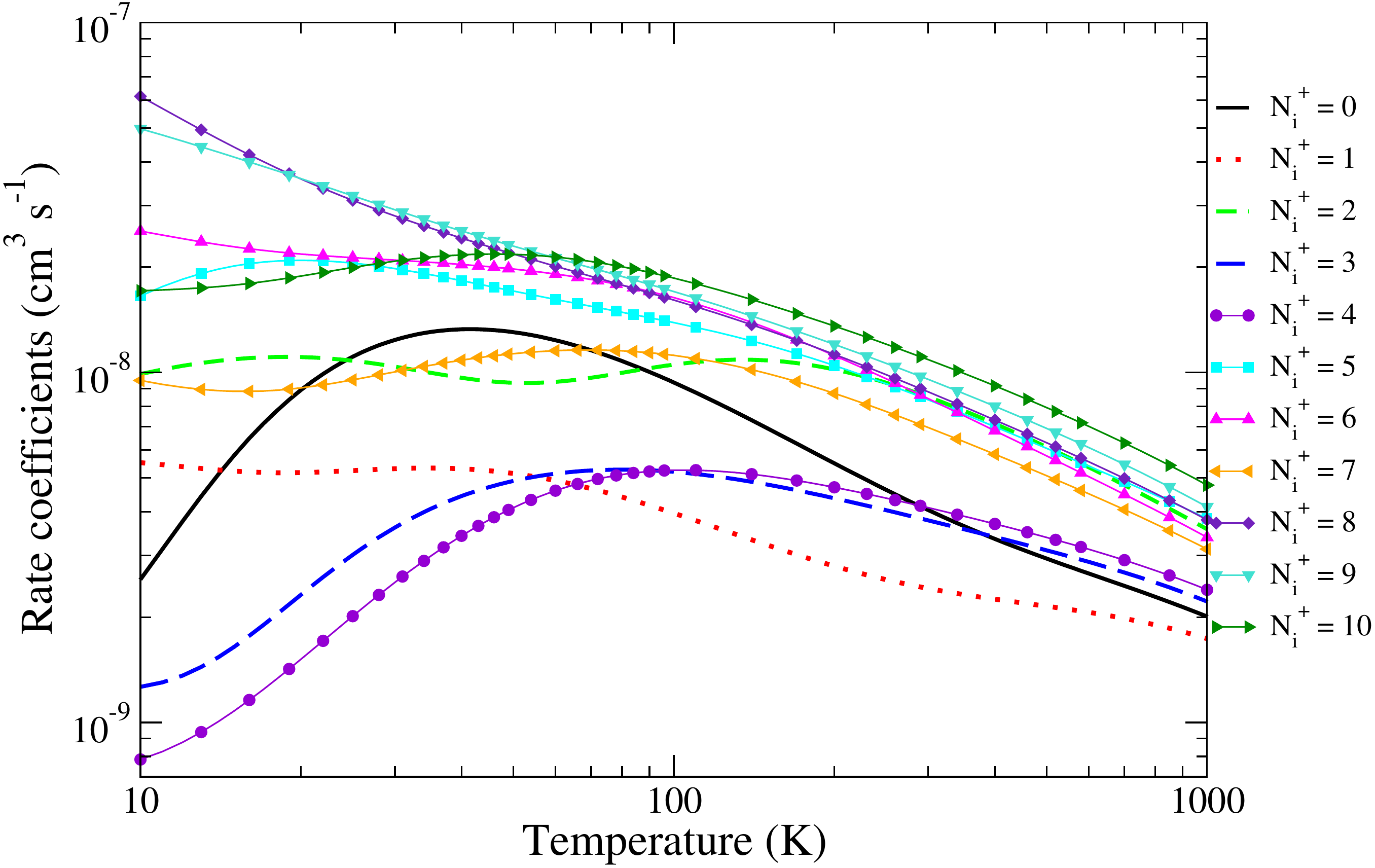}
    \caption{Maxwell rate coefficients for the dissociative recombination of D$_{2}^{+}$ (X $^{2}\Sigma_{g}^{+},v_{i}^{+}=0)$ as a function of its initial rotational levels, $N_{i}^{+}$, varying from $0$ to $10$.}
    \label{fig:r_DR_iso}
\end{figure}

\begin{figure}
		\includegraphics[width=\columnwidth]{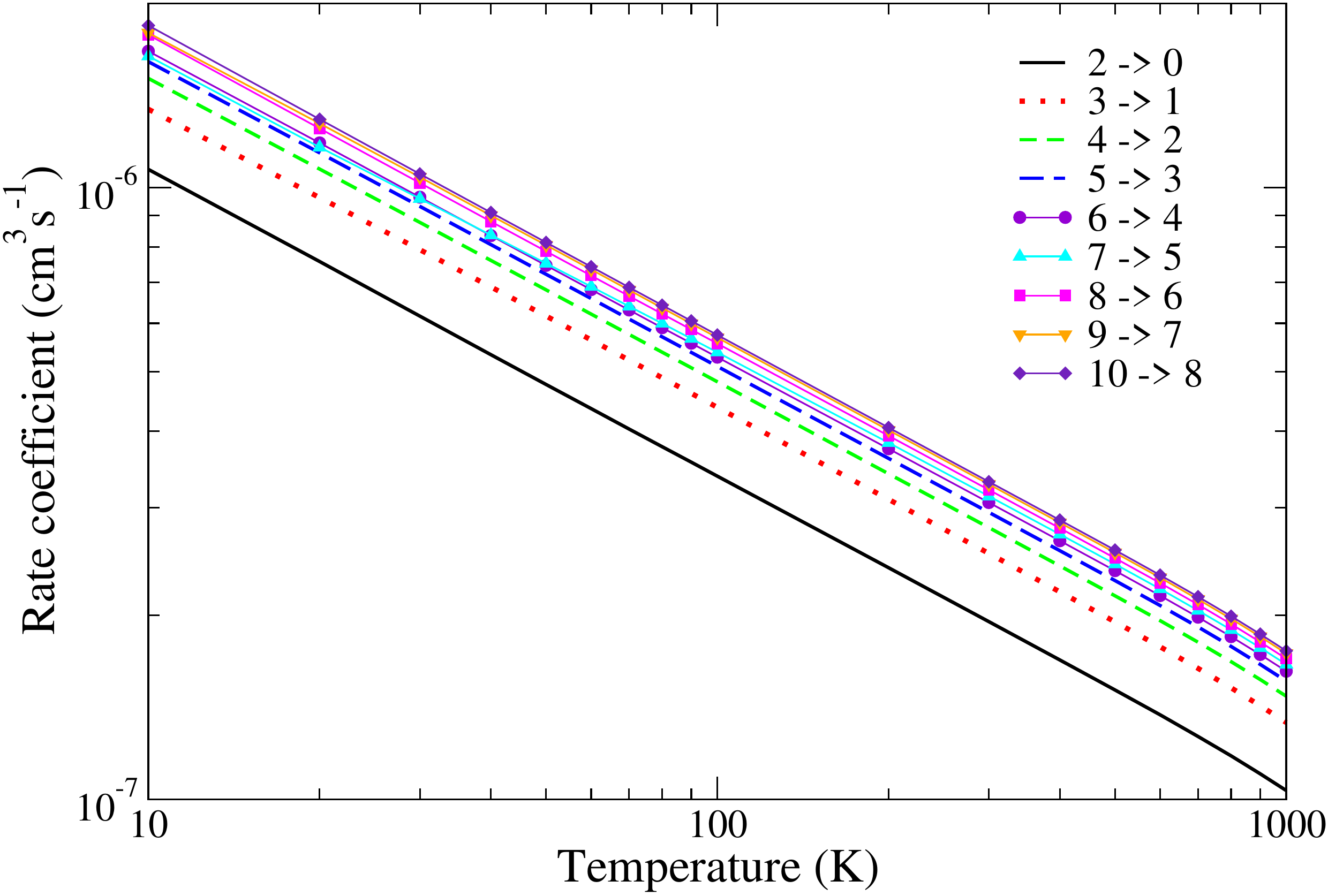}
    \caption{
    Maxwell rate coefficients for the rotational de-excitation $N_{i}^{+} \rightarrow N_f^{+}=N_{i}^{+}-2$, with $N_{i}^{+}=2$ to $10$ of the vibrationally relaxed $(v_i^+=0)$ D$_{2}^{+}$ (X $^{2}\Sigma_{g}^{+})$.
    }
    \label{fig:r_RdE_iso}
\end{figure}

\begin{figure}
		\includegraphics[width=\columnwidth]{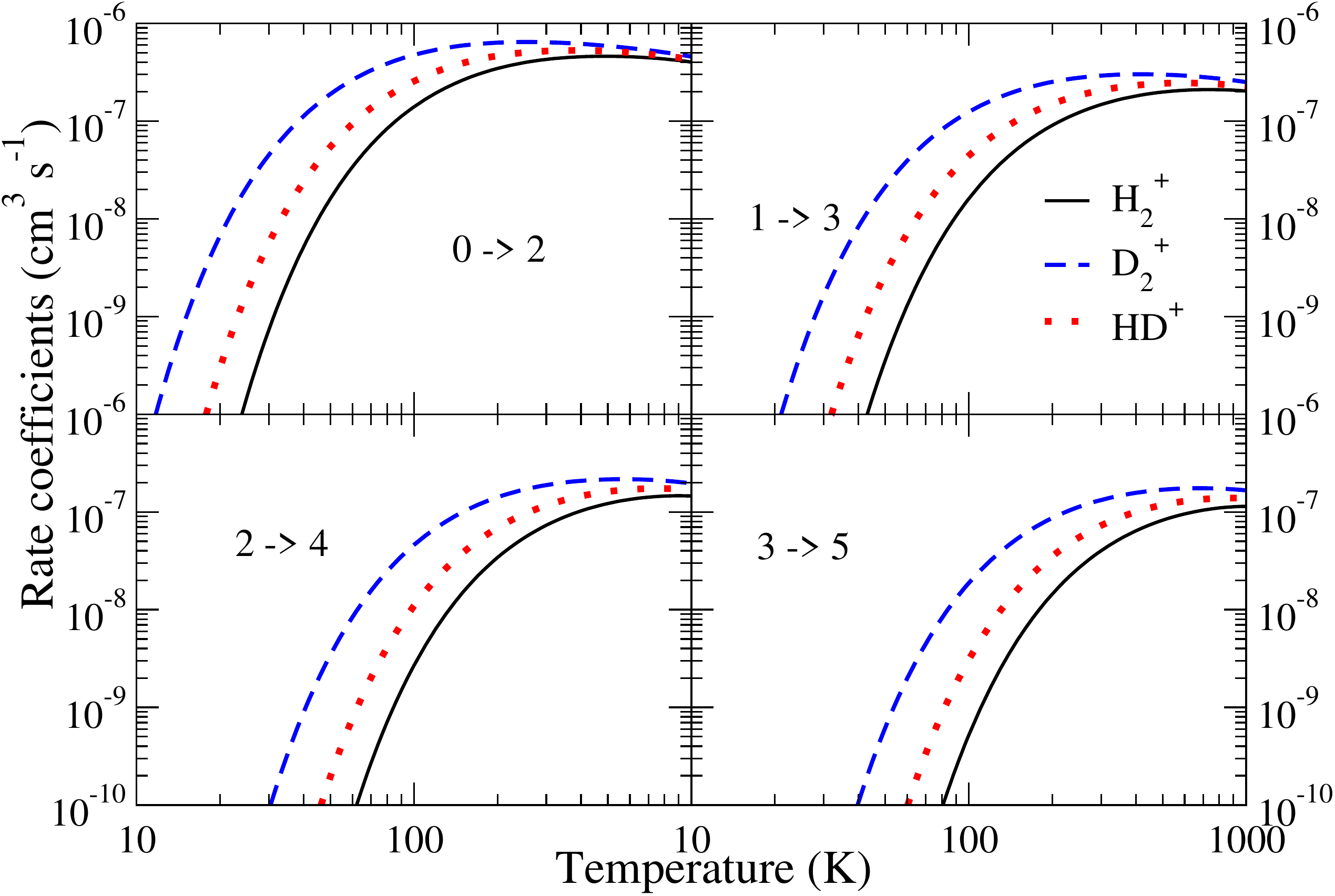}
   \caption{ Isotopic effects in rotational excitation: rate coefficients for 
$N_i^+ \rightarrow N_f^{+}=N_i^{+}+2$ transitions, $N_{i}^{+}=0$ to $3$,
for the vibrationally
relaxed $X^{2}\Sigma_{g}^{+}$ H$_{2}^{+}$, HD$^{+}$   and D$_{2}^{+}$ systems.}
    \label{fig:compa_H2_D2}
\end{figure}
\begin{figure}
		\includegraphics[width=\columnwidth]{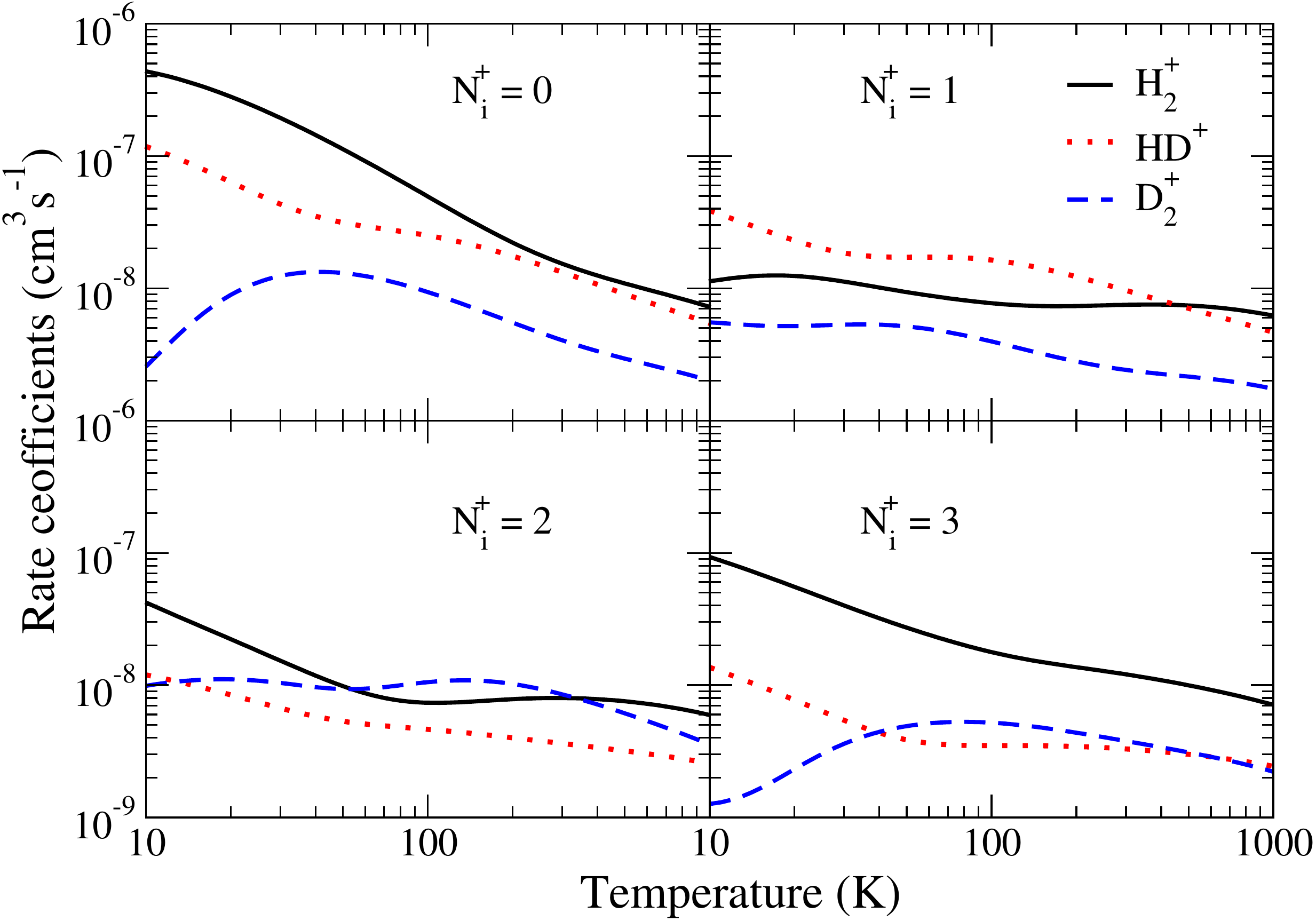}
   \caption{ Isotopic effects in dissociative recombination: rate coefficients for $N_{i}^{+}=0$ to $3$, for the vibrationally relaxed $X^{2}\Sigma_{g}^{+}$ H$_{2}^{+}$, HD$^{+}$   and D$_{2}^{+}$ systems.}
    \label{fig:compa_DR}
\end{figure}

\begin{figure}
		\includegraphics[width=\columnwidth]{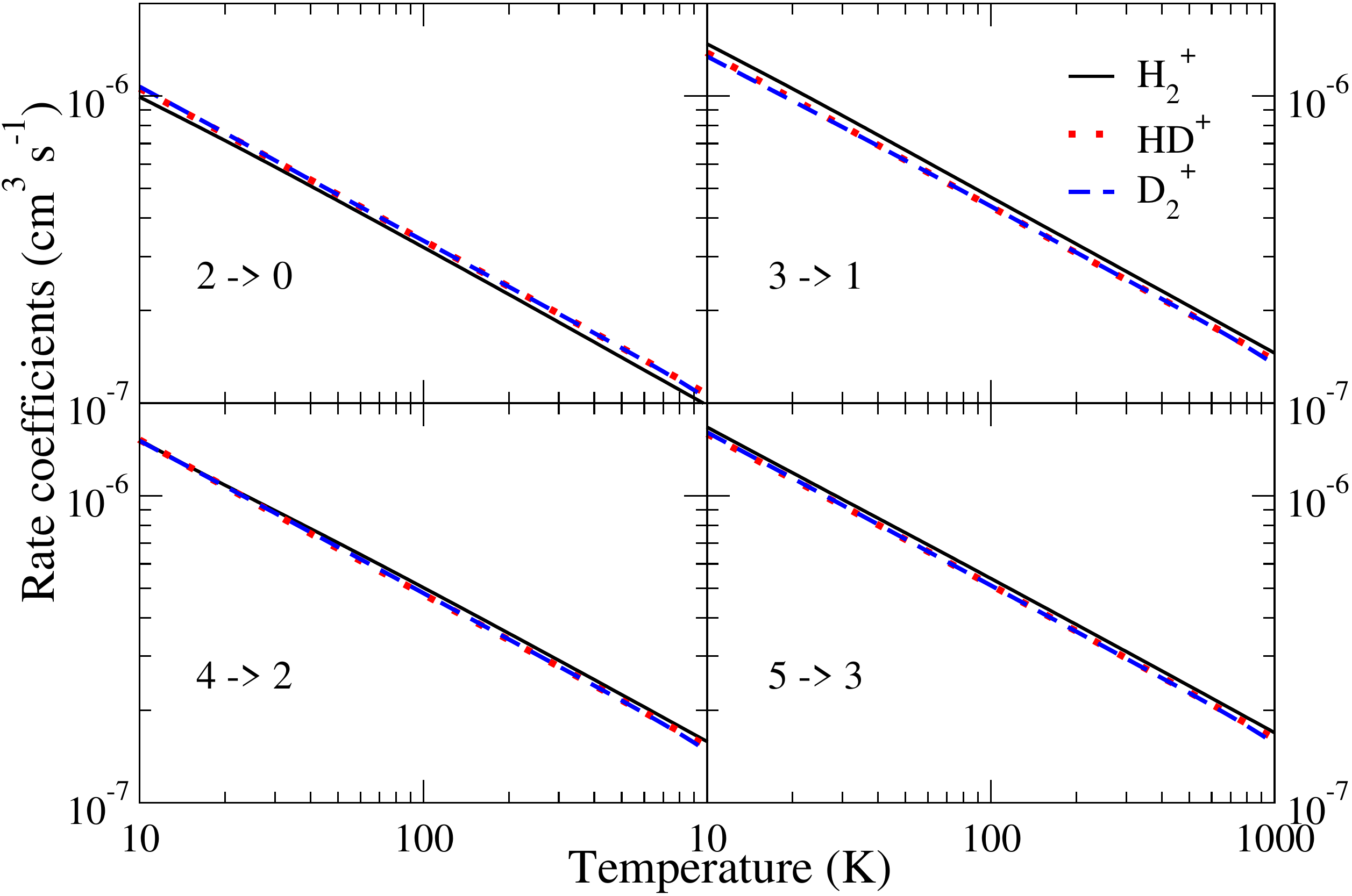}
   \caption{ 
   Isotopic effects in rotational de-excitation: rate coefficients for $N_i^+ \rightarrow N_f^{+}=N_i^{+}-2$ transitions, $N_{i}^{+}=2$ to $5$,
for the vibrationally relaxed $X^{2}\Sigma_{g}^{+}$ H$_{2}^{+}$, HD$^{+}$   and D$_{2}^{+}$ systems.
}
    \label{fig:compa_RdE}
\end{figure}

\begin{table}
	\centering
	\caption{Comparison of Maxwell rate coefficients (in cm$^{3}$ s$^{-1}$) for $N_{i}^{+}\rightarrow N_{f}^{+}=N_{i}^{+}+2$ rotational excitations of H$_{2}^{+}$, HD$^{+}$ and D$_{2}^{+}$ by collision with electrons at room temperature ($T=300$ K). }
	\label{tab:IC_H2_HD_D2}
	\begin{tabular}{lccc} 
		\hline
		 $N_{i}^{+}$ & H$_{2}^{+}$ & HD$^{+}$& D$_{2}^{+}$ \\
		\hline
		0 & 4.27473E-7 & 5.20549E-7 & 6.40280E-7 \\
		1 & 1.47448E-7 & 2.06564E-7 & 2.91847E-7 \\
		2 & 7.30450E-8 & 1.16444E-7 & 1.88170E-7 \\
		3 & 4.0060E-8   & 7.26357E-8 & 1.32346E-7 \\
		4 & 2.30159E-8 & 4.70647E-8 & 9.65611E-8 \\
		5 & 1.36503E-8 & 3.12933E-8 & 7.16222E-8 \\
		6 & 8.54116E-9 & 2.15608E-8 & 5.46144E-8 \\
		7 & 5.56954E-9 & 1.49720E-8 & 4.19320E-8 \\
		8 & 3.69408E-9 & 1.04648E-8 & 3.26509E-8 \\
		9 & 2.45815E-9 & 8.02520E-9 & 2.57048E-8 \\
		10 & 1.72925E-9 & 5.91376E-9 & 2.03948E-8 \\
		\hline
	\end{tabular}
\end{table}

Contrary to this, the DR Maxwell rate coefficients shows a more complicated behaviour as function of electron temperature and target initial rotational quantum number, as one can see in Figure \ref{fig:r_DR_iso}. The rate coefficients obtained for the different initial rotational levels vary between one and two orders of magnitude as function of the electron temperature. 
\begin{table*}
	\centering
	\caption{Fitting parameters for the formula (\ref{eq:fit-DR}), corresponding to the rate coefficients for dissociative recombination of vibrationally relaxed D$_{2}^{+}$ on its lowest 11 rotational levels ($N_{i}^{+}=0-10$, $v_{i}^{+}=0$) with electrons of temperature in the range $10-1000$ K.}	
\label{tab:rate_DR_fitt}
	\begin{tabular}{lccccccc} 
		\hline
		 $N_{i}^{+}$ & a$_0$(cm$^{3}$ s$^{-1}$) & a$_1$ & a$_2$(K)& a$_3$(cm$^{3}$ s$^{-1}$) & a$_4$ & a$_5$(K)&RMS\\
		\hline
0	&	1.61858E-9	&	-0.026610	&	1.35036	&	2.97304E-9 & -1.38979 & 57.3832 & 0.00841	\\
1	&	2.41400E-9	&	-0.281201	&	1.25069	&	1.15502E-10 & -2.87398 & 137.585 & 0.01566	\\
2	&	3.58811E-9	&	-0.609918	&	10.6046	&	1.24094E-8 & -1.41135 & 264.577 & 0.00977	\\
3	&	6.49161E-10	&	0.092753	&	-8.57227	&	3.73310E-9 & -0.674586 & 54.0504 & 0.00723	\\
4	&	1.43940E-9 	&      -0.07285 & 	8.76012 	&	3.71294E-9 & -0.873084 & 93.3021 & 0.01135 \\
5	&	5.20107E-9	&	-0.804386	&	15.7873	&	6.03106E-9 & -0.885171 & 164.260 & 0.00710	\\
6	&	3.40860E-9	&	-0.737167	&	5.06339	&	6.55842E-9 & -0.945548 & 72.6807 & 0.00601	\\
7	&	1.56377E-9	&	-0.662548	&	4.60987	&	6.96067E-9 & -0.800087 & 73.4612 & 0.00647	\\
8	&	5.08388E-9	&	-0.677241	&	-1.88673	&	4.75096E-9 & -0.792344 & 78.2720 & 0.00783	\\
9	&	7.44652E-9	&	-0.615038	&	1.93963	&	3.35883E-9 & -1.079090 & 140.250 & 0.00843	\\
10	&	4.02898E-9	&	-0.531062	&	3.93537	&	8.19758E-9 & -0.827386 & 54.8747 & 0.00878	\\
		\hline
	\end{tabular}
\end{table*}
Up to $200$ K, the most populated target rotational level (at local thermal equilibrium) gives the major contributions to the rate coefficient. Above this temperature the higher rotational quantum numbers become more and more important. Depending on the rotational quantum number of the initial state of the target, the rates show various temperature dependencies, from the smoothly decreasing  behaviour to more general functions showing at least one maximum - see Figure 1. While for low rotational quantum numbers the DR rate coefficients are exceeded by RE  already at electron temperatures smaller then 50 K, for $N_{i}^{+}=10$ target this takes place at $T\sim300$ K only. 

The thermal rate coefficients obtained for the $N_i^+ \rightarrow N_f^{+}=N_i^{+}-2$ rotational de-excitations of D$_2^+$ for $N_{i}^{+}=2$ to $10$ and $v_i^+=v_f^+=0$ are given in Figure~\ref{fig:r_RdE_iso}. The magnitude of rate coefficients increase with the initial rotational quantum numbers. It is also notable that below $1000$ K, they are larger then those of the dissociative recombination. 

In Figure~\ref{fig:compa_H2_D2} we compare the thermal rate coefficients for the rotational excitation from the lowest four rotational levels of H$_{2}^{+}$ (continuous black line), HD$^+$ (dotted red line) and D$_{2}^{+}$ (dashed blue line). The isotopic effects mainly due to the sharp thresholds are notable (notice the use of the logarithmic scale). We found that heavier the isotopologue, the larger the rate coefficient. This effect is quantified in table \ref{tab:IC_H2_HD_D2} for the first eleven initial rotational levels of the targets for collisions taking place at $T=300$ K temperature.

 Figure~\ref{fig:compa_DR}  presents the isotopic effects obtained for DR for the lowest four rotational quantum numbers of the vibrationally relaxed target. Due to the very different dependence of the DR process on the initial and final channels with respect to the rotational excitation, we obtain different isotopic effects compared to those of Figure~\ref{fig:compa_H2_D2}. Except for $N_i^+=2$ case where the rate for D$_2^+$ is very close to that of H$_2^+$ and exceeds the one of HD$^+$ and for $N_i^+=3$ for $T=40-400$ K, the rates obtained for the heaviest isotopologue are the smallest among all.

Similarly to DR and RE, in figure~\ref{fig:compa_RdE} we have compared the RdE rate coefficients for the three isotopologues for the lowest four $\Delta N=-2$ transitions. The dependence of the RdE on the initial and final channels and the lack of threshold effects, in contrary to RE, lead to the a slight isotopic effect.

 In order to facilitate the use of our recombination, excitation and de-excitation rate coefficients for kinetic modelling, we have fitted their temperature dependence by using Arrhenius-type formulas.

 For the DR we used:
 
\begin{equation}
\alpha(T) =a_0\left(\frac{T}{300}\right)^{a_1}e^{-\frac{a_2}{T}}+a_3\left(\frac{T}{300}\right)^{a_4}e^{-\frac{a_5}{T}}, 
 \label{eq:fit-DR}
\end{equation}
and for the rotational transitions:
\begin{equation}
\alpha(T) =a_0\left(\frac{T}{300}\right)^{a_1}e^{-\frac{a_2}{T}}, 
\label{eq:fit-RE}
\end{equation}

\noindent where $T$ is in Kelvin and $\alpha$ in cm$^{3}$s$^{-1}$. The fitting parameters for the DR of the lowest 11 rotational levels of the target, the 9 rotational de-excitation and 11 rotational excitation rate coefficients are summarized in tables~\ref{tab:rate_DR_fitt}, \ref{tab:rate_RDE_fitt} and \ref{tab:rate_RE_fitt}. For all the processes, the fitted values reproduce well our MQDT rate coefficients, according to the RMS values given in the forth column of each table in the whole temperature range from $10 < T< 1000$ K.  

In addition to the consecutive transitions we have also calculated the rate coefficients for rotational transitions with $\Delta N^{+}=4$. We have found that they are about two orders of magnitude smaller than those obtained for $\Delta N^{+}=2$ for the same initial rotational quantum number.  Consequently,  we have omitted them from the present paper. And finally, besides the symmetry allowed $\Delta N^+=2,4$ rotational transitions of the H$_2^+$ and D$_2^+$ cations one has to mention that the $\Delta N^+=1$ rotational transitions in HD$^+$ are significant due to the existing permanent dipole moment ($\mu=0.85$ Debye)~\cite{shafir}. The theoretical treatment of these transitions due to the inexistent "gerade-ungerade" couplings is a serious challenge. 

\begin{table}
	\centering
	\caption{Fitting parameters for the formula (\ref{eq:fit-RE}),  corresponding to the rate coefficients for rotational de-excitation 
	$N_{i}^{+}\rightarrow N_{f}^{+}=N_{i}^{+}-2$
	of vibrationally relaxed ($v_{i}^{+}=0$) D$_{2}^{+}$ on its  rotational levels $N_{i}^{+}=2-10$ with electrons of temperature in the range $10-1000$ K.}	
	\label{tab:rate_RDE_fitt}
	\begin{tabular}{lcccc} 
		\hline
		 $N_{i}^{+}$ & a$_0$(cm$^{3}$ s$^{-1}$) & a$_1$ & a$_2$(K) & RMS\\
		\hline
2	&	1.94131E-7	&	-0.502252	&	0.006921	&	0.00745	\\
3	&	2.51942E-7	&	-0.503619	&	0.401653	&	0.00813	\\
4	&	2.77168E-7	&	-0.503306	&	0.185143	&	0.00724	\\
5	&	2.93427E-7	&	-0.503931	&	0.145077	&	0.00697	\\
6	&	3.03986E-7	&	-0.502306	&	0.044583	&	0.00208	\\
7	&	3.12123E-7	&	-0.490849	&	0.120819	&	0.01280	\\
8	&	3.19233E-7	&	-0.503000	&	-0.053667	&	0.00742	\\
9	&	3.26264E-7	&	-0.506715	&	0.214215	&	0.00178	\\
10	&	3.29173E-7	&	-0.504694	&	-0.042102	&	0.00713	\\
		\hline
	\end{tabular}
\end{table}

\begin{table}
	\centering
	\caption{Fitting parameters for the formula (\ref{eq:fit-RE}),  corresponding to the rate coefficients for rotational excitation 
	$N_{i}^{+}\rightarrow N_{f}^{+}=N_{i}^{+}+2$
	of vibrationally relaxed ($v_{i}^{+}=0$) D$_{2}^{+}$ on its  rotational levels $N_{i}^{+}=0-10$ with electrons of temperature between $T_{\text{min}}$ and $1000$ K, where $T_{\text{min}}$ is the temperature below which the rate coefficient is lower than $10^{-14}$ cm$^3$s$^{-1}$.}
	
	\label{tab:rate_RE_fitt}
	\begin{tabular}{lccccc} 
		\hline
		 $N_{i}^{+}$ & $T_{\text{min}}$(K) & a$_0$(cm$^{3}$ s$^{-1}$) & a$_1$ & a$_2$(K) & RMS\\
		\hline
0	& 10 &	9.80558E-7	&	-0.510696 	&	128.209	&	0.02427	\\
1	& 11 & 	5.98558E-7	&	-0.528565 	&	215.309	&	0.06229	\\
2	& 16 & 	5.16852E-7	&	-0.541313 	&	302.347	&	0.09660	\\
3	& 20 & 	4.88827E-7	&	-0.560415		&	390.647	&	0.14360	\\
4	& 25 & 	4.80191E-7	&	-0.582399		&	479.527	&	0.18399	\\
5	& 29 & 	4.77666E-7	&	-0.603300		&	567.434	&	0.21397	\\
6	& 33 &	4.81761E-7	&	-0.624484		&	651.362	&	0.23555	\\
7	& 37 & 	4.87513E-7	&	-0.645269		&	734.352	&	0.24984	\\
8	& 41 & 	4.96209E-7	&	-0.666460		&	814.971	&	0.25815	\\
9	& 45 & 	5.07156E-7	&	-0.687319		&	893.630	&	0.26111	\\
10	& 49 &	5.17958E-7	&	-0.709294		&	969.819	&	0.25649	\\
		\hline
	\end{tabular}
\end{table}

\section{Conclusions}{\label{sec:conclusions}}

In the framework of the stepwise multichannel quantum defect theory we have calculated cross sections between $0.01$ meV  and  $0.3$ eV, and consequently thermal rate coefficients between $10$ and $1000$ K, for dissociative recombination and rotational excitation/de-excitation of electrons with D$_{2}^{+}(X^{2}\Sigma_{g}^{+})$ ions for their lowest $11$ rotational levels and in their ground vibrational level $v_{i}^{+}=v_{f}^{+}=0$. 

In our model we have accounted for all relevant electronic states and symmetries of the cation target, for all relevant rotational and vibronic electronic couplings, by taking into account the quantum interference among the direct and indirect mechanisms.

The obtained rate coefficients show strong dependence on the initial rotational state of the molecular target.

We have compared the present dissociative recombination and rotational excitation/de-excitation coefficients obtained for D$_{2}^{+}$ with similar rate coefficients previously calculated for H$_{2}^{+}$ and HD$^{+}$ isotopologues. They crucially depend on fine balance between the initial and final channels and threshold effects. For rotational excitation we observe that heavier the cation, larger the rate coefficient, while for de-excitation we get only a slight isotopic effect. The strongest initial/final channel dependence can be observed for the dissociative recombination. The obtained isotopic differences clearly put in evidence the importance of the present results especially for kinetic modelling of the environments where deuterated species are present.

These results complement  significantly the recent investigations on the other main competing destructive channels of H$_2^+$, HD$^+$ and D$_2^+$ via their reactions with H$_2$, HD and D$_2$ that produce the H$_3^+$, H$_2$D$^+$ , D$_2$H$^+$ and D$_3^+$ triatomic ions \citep{merkt:22}, allowing to remove significant uncertainties of previous studies.

The numerical data, ready to be used in the kinetic modelling in astrochemistry and cold plasma physics will be available upon request. 

\section*{Acknowledgements}
%%%%%%%%%%%%%%%%%%%%%%%%%%%%%%%%%%%%%%%%%%%%%%%%%%
The authors acknowledge support from F\'ed\'eration de Recherche Fusion par Confinement Magn\'etique (CNRS and CEA), La R\'egion Normandie, FEDER, and LabEx EMC3 via the projects PTOLEMEE, Bioengine COMUE Normandie Universit\'e, the Institute for Energy, Propulsion and Environment (FR-IEPE), the European Union via COST (European Cooperation in Science and Technology) actions TUMIEE (CA17126), MW-Gaia (CA18104) and MD-GAS (CA18212) and from the l'Agence Universitaire de la Francophonie en Europe Centrale et Orientale (AUF ECO) via the project CE/MB/045/2021 CiCaM -- ITER.
The authors are indebted to Agence Nationale de la Recherche (ANR) via the project MONA. This work was supported by the Programme National "Physique et Chimie du Milieu Interstellaire" (PCMI) of CNRS/INSU with INC/INP co-funded by CEA and CNES. J.Zs.M. thanks the financial support of the National Research, Development and Innovation Fund of Hungary, under the FK 19 funding scheme with project no. FK 132989.

%\section*{Data availability}
%The data underlying this article will be shared on reasonable request to the corresponding author.

\end{document}